\documentclass[twocolumn,showpacs,preprintnumbers,amsmath,amssymb]{revtex4}

\usepackage{graphicx}
\usepackage{dcolumn}
\usepackage{bm}

\begin{document}

\title{Low Energy Transfer Cross-Section for Borromean Halo Nuclei}
\author{W.H.Z. C\'{a}rdenas$^1$, L.F.~Canto$^2$ and M.S.~Hussein$^1$ \\}

\address{$^{1}$Instituto de F\'{\i}sica, Universidade de S\~{a}o Paulo, C.P. 66318, 05389-970 S\~{a}o Paulo, Brazil\\
$^2$Instituto de F\'{\i}sica, Universidade Federal do Rio de Janeiro, C.P. 68528, 21945-970 Rio de Janeiro, Brazil,\\}

\begin{abstract}
 We describe a schematic coupled channels transfer
calculation for the reaction $^6$He + $^{238}$U at near-barrier energies. We also present a simple semiclassical DWBA calculation of the two neutron transfer.
Both calculations are meant to supply under conditions at which the transfer
cross section becomes much larger than the complete fusion one at sub-barrier
energies. It seems that a feasible mechanism is the incoherent contributions of two or more processes with quite different Q-values.  
\end{abstract}

\date{\today}

\pacs{Valid PACS appear here}
\maketitle
Cross-sections for fusion reactions with neutron halo nuclei in the energy region below the Coulomb barrier are necessary for calculating the thermonuclear reaction rates in massive stars. In addition, such reactions provide useful information about the shape of the nuclear potential on the inner side of interaction barrier. Furthermore, interpretation of such cross-sections may bring possible gain information on the influence of the distribution of nuclear matter and the nuclear reaction dynamics, especially for those energies where penetrability effects are important \cite{Ca05}. 

Recently, nuclear reactions involving the neutron-rich nucleus $^6$He have
attracted considerable attention. In particular, very interesting experimental data on the fusion of He isotopes with $^{238}$U have been obtained. These data show no enhancement of the $^6$He+$^{238}$U fusion cross section, but a very high transfer cross section has been observed \cite{Tr00,Ra04}. The physical process leading to this result has not yet been established. The natural candidates are the coupling with the breakup and transfer channels. However, understanding the effect of the neutron halo on fusion has been controversial, since the weakly bound neutrons in $^6$He are expected to influence the fusion cross section in two ways. Firstly, by the static effect of barrier lowering due to the existence of a halo. Secondly through the coupling with the breakup channel. Also, in neutron halo nuclei reaction, the neutron transfer cross section should play an important role in sub-barrier fusion of heavy nuclei, due to the small binding energy of neutrons halo and the positive $Q$-value.

Furthermore, reactions with stable nuclei, at energies below Coulomb barrier are caracterized by large enhancements in the fusion  cross section with respect to calculations based on one dimensional barrier penetration models. It has been quite well understood that these enhancements are due to the coupling to different degres of freedom acting on the tunneling process, mainly static deformations and surface vibrations of nuclei. The role of transfer channels is however still unclear.

In this short paper we describe a schematic coupled channels transfer
calculation for the reaction $^6$He + $^{238}$U at near-barrier energies. We also present a simple semiclassical DWBA calculation of the two neutron transfer.
Both calculations are meant to supply under conditions under which the transfer
cross section becomes much larger than the complete fusion one at sub-barrier
energies. In fact, such a situation seem to prevail for light systems  \cite{Ca03}. For the purpose of completeness, we also calculate the fusion cross section and demonstrate that there is little difference when compared to that of the system $^4$He + $^{238}$U.

In the calculation to follow we take as optical potential the single folding one given by the integral
\begin{equation}
V_N(r)=\int v_{n-A_2}({\bold r}-{\bold r}')\rho({\bold r}')d^3{\bold r}'.
\end{equation}
Above, $v_{n-A_2}$ is an appropriate nucleon-target interaction and $\rho({\bold r}')$ is the projectile's density. The full optical potencial is
\begin{equation}
U(r)=V_N(r)-iW(r)+V_C(r),
\label{op}
\end{equation}
where $V_C(r)$ is the Coulomb potential.

A good description of the fusion cross section for collision of the stable isotope $^4$He + $^{238}$U is obtained when we use the real part of the nucleon-target interaction of Madland and Young \cite{Ma78} and a Gaussain form for the projectile's density. For the imaginary part we take a Woods-Saxon parametrization with $W_0 = 50$ MeV, $r_i=1.0$ fm and $a_i=0.10$ fm.

For the real part of the interaction of the $^6$He + $^{238}$U system we merely use a different density profile to take into account the two-neutron halo in a realistic parametrization given by the symmetrized Fermi distribution of \cite{Al97},
\begin{widetext}
\begin{equation}
\rho_{^6He}(r)=\rho_0\left[ \left( 1+\exp\left( \frac{r-R}{a}\right)\right)^{-1}+\left( 1+\exp\left( \frac{-r-R}{a}\right)\right)^{-1}-1\right]
\end{equation}
\end{widetext}
with
\begin{equation}
\rho_0=\frac{3A}{4\pi R^3}\left[ 1-\left(\frac{\pi a}{R}\right)^2\right]^{-1},
\end{equation}
$A=6$, $R=1.23A^{1/3}$ fm and $a=0.57$ fm.

The fusion cross section may be expressed in terms of the optical model transmission factor as follows,
\begin{equation}
\sigma_F = \frac{\pi}{k^2}\sum_{l=0} (2l+1)T^F_l,
\label{sigfus}
\end{equation}
where $k=\sqrt{2\mu E/\hbar^2}$ and the transmission coefficient is given by
\begin{equation}
T^F_l=\frac{4k}{E}\int^\infty_0 dr W^{opt}(r)|u_l(k,r)|^2,
\label{tlf}
\end{equation}
with, $W^{opt}(r)$ is the imaginary part of the optical potential and $u_l(k,r)$ are solutions of the radial equation.

\begin{figure*}
\includegraphics[width=13.cm,height=12.cm]{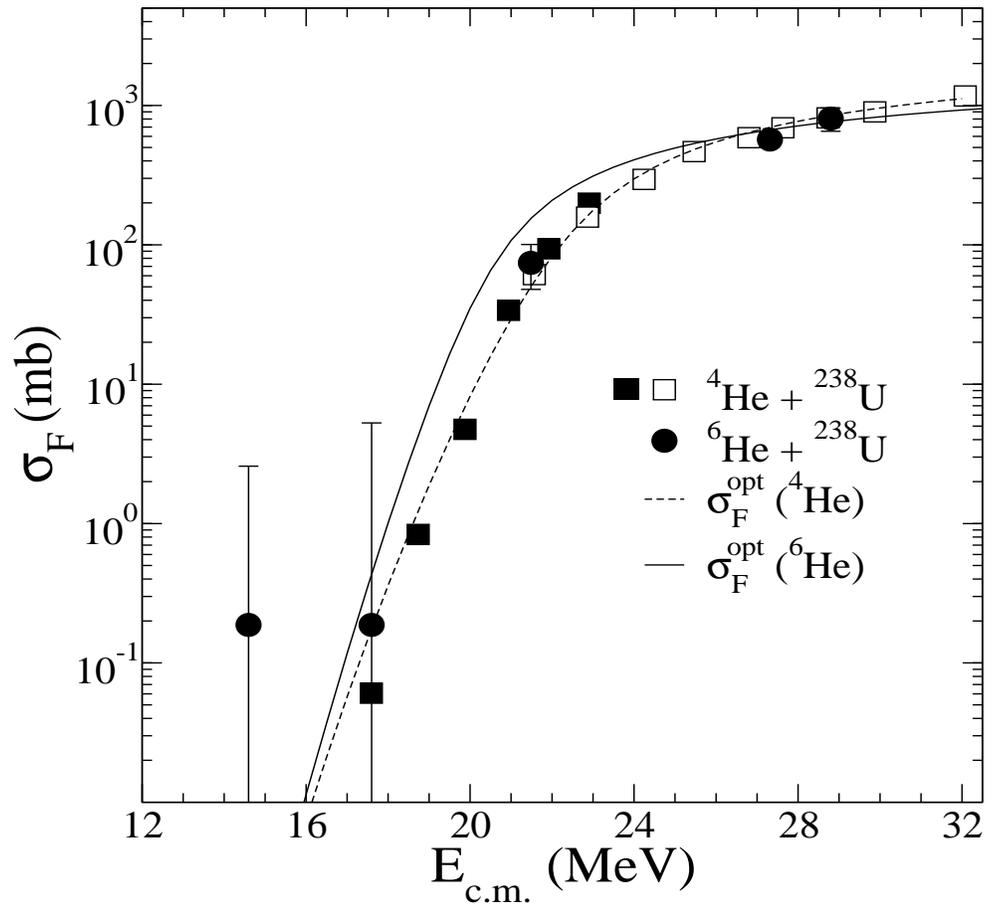}%
\caption{\label{fig1} A figure caption.}
\end{figure*}

We have calculated, within a one-channel optical model, the fusion of $^{4,6}$He with $^{238}$U using the above interaction. The results are shown in figure \ref{fig1}. Clearly very little difference, owing to the static halo effect in $^6$He, is found in comparison to $^4$He. The result for $^6$He fusion was multiplied by a factor 0.6, to account for elastic breakup and other processes as was found in Ref. \cite{Da99,Hi02} for the system $^9$Be + $^{208}$Pb. In fact the elastic breakup of $^6$He leading to two flying out neutrons clearly does not contribute to the fission events, which are attributed to 2n removal or transfer. This is borne out by the data as well. We turn to the two-neutron removal cross section.

Within the WKB approximation we may take as the breakup survival probability, $P_l^{surv}$ \cite{Ca02}
\begin{equation}
P_l^{surv}=\exp\left[ \beta^I_l\right]
\end{equation}
where 
\begin{equation}
\beta^I_l=-\frac{4}{\hbar}\int_{r_0(l)}^\infty\frac{W^{n}(R)}{v_l(r)}dr.
\end{equation}
The quantities, $v_l(r)$ and $r_0(l)$ are the local radial velocity along a classical trayectory with momentum $\hbar l$ and the closest distance of approach, respectively, for the relative motion in the nuclear, Coulomb and centrifugal potentials, and $W^{n}$ the neutron absortion potential. Thus, the transfer cross section is obttained by considering that

\begin{equation}
T_l^{T}=1-P_l^{surv}.
\end{equation}

In our calculations, in order to estimate $P_l^{surv}$ we have considered pure Rutherford trajectories, neglecting the nuclear potential diffractive effects. For the absorptive potential we use the original  Madland-Young, imaginary potential \cite{Ma78} which describes very well neutron scattering from actinide nuclei at $E_n < 10$MeV. It is given by
\begin{widetext}
\begin{eqnarray}
W^n(R)&=&-4a_iW_0 f'_I(R)\nonumber\\
f_I(R)&=&\left[ 1+\exp\left(\frac{R-R_I}{a_I}\right)\right]^{-1}\nonumber\\
W_0&=&9.265-12.666\left[ \frac{N-Z}{A}\right]-0.232E_{Lab} + 0.03318E_{Lab}^2\\
R_I&=&1.256A_T^{1/3}\nonumber\\
a_I&=&0.553 + 0.0144E_{Lab}\nonumber
\end{eqnarray}
\end{widetext}

Let us consider the collision of $^6$He, treated within the dineutron approximation, with a heavy target. The imaginary part can be written
\begin{equation}
W^n(r,x)=W^n(|{\bf R|})
\label{wrx}
\end{equation}
where ${\bf R}= {\bf r} + 2{\bf x}/3$, here ${\bf r}$ is the vector joining the centers of mass of projectile and target and ${\bf x}$ is the vector from the $^4$He core to the dineutron.

Thus, from Eqs. (\ref{sigfus}-\ref{wrx}) 2n-removal cross section is given by
\begin{equation}
\sigma_{-2n}(x)=\frac{\pi}{k^2}\sum_{l=0}(2l+1)T^{2n}_l(x)
\end{equation}
where
\begin{equation}
T^{2n}_l(x)=1-\exp\left[ \beta^I_l(x)\right].
\end{equation}

The expected value of the 2n-removal cross section is given by

\begin{equation}
\bar{\sigma}_{-2n}=\langle \phi_0| \sigma_T|\phi_0\rangle
\label{sigtra}
\end{equation}
where $\phi_0$ describes the ground state of the projectile in its rest frame, and is a function of the relative coordinates of de 2n-halo and the core,

\begin{equation}
\phi_0({\bf x})=(2\pi\alpha)^{-1/2}\frac{e^{-x/\alpha}}{x},\qquad \alpha=\frac{\hbar}{\sqrt{2B_{2n}\mu_{(^6He)}}}
\label{phix}
\end{equation}
$B_{2n}$ (= 0.973 MeV)  is the dineutron binding energy in $^6$He and $\mu_{(^6He)}=\frac{4}{3}m_0$, is the reduced mass of the $^{6}$He system being $m_0$ the nucleon mass. From equation (\ref{phix}), we obtain the root mean square radius $r_{rms}=2.84$ fm.

\begin{figure*}
\includegraphics[width=13.5cm,height=12.cm]{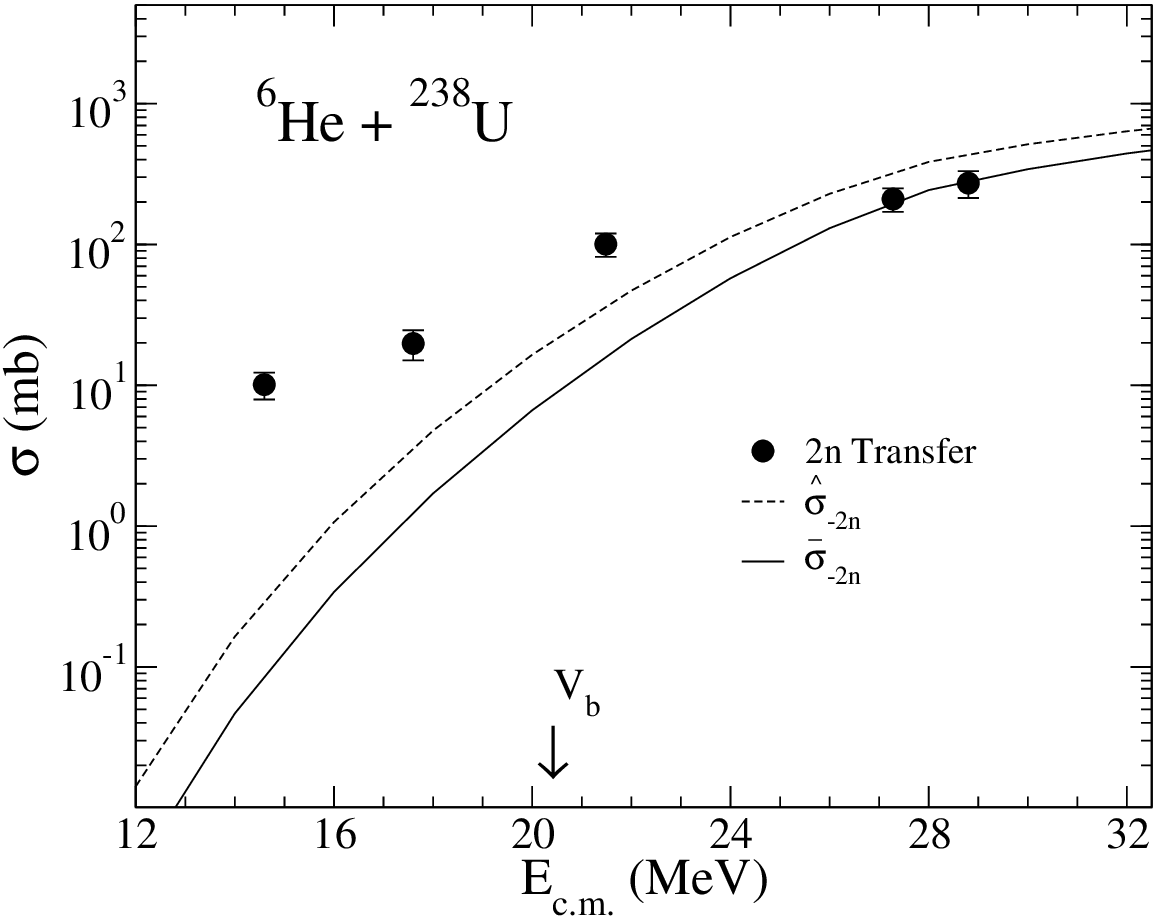}
\caption{\label{fig2} A figure caption.}
\end{figure*}

In Figure \ref{fig1}, we shall compare $\bar{\sigma}_{-2n}$, Eq. (\ref{sigtra}), with the expression
\begin{equation}
\hat{\sigma}_{-2n}=\frac{\pi}{k^2}\sum_{l=0}(2l+1)\hat{T}^{2n}_l(x)
\label{sigtra1}
\end{equation}
where
\begin{equation}
\hat{T}^{2n}_l(x)=1-\exp\left[ \langle\phi_0|\beta^I_l|\phi_0\rangle\right].
\end{equation}
From Peierls' inequality \cite{Pe38}
\begin{equation}
\langle \exp F \rangle\ \geqslant \exp\langle F \rangle,
\end{equation}
we have
\begin{equation}
{ \bar \sigma} \leqslant {\hat \sigma},
\end{equation}
as seen in figure \ref{fig2}, for the range of energies shown here. This difference is mainly due to the great spatial distribution of nucleons in the halo, leading to an  uncertainty in the location of the nucleons.

Both calculations of the 2n removal cross section miss completely the data. This calls for a different approach. Before we turn to our next attempt in understanding the nature of the sub-barrier transfer data, we calculate the amount of angular momentum transferred in the complete fusion ( to the compound nucleus $^{244}$Pu) and in the 2n transfer process (to the isotope $^{240}$U). The formulae we use for this purpose were derived in Refs. \cite{Ca05,Ca98}, and we give only  the results here. At below-barrier energies the complete fusion transfers about 4 units of $\hbar$ to $^{244}$Pu while the two neutrons transfer just one unit
of $\hbar$ to $^{240}$U. These two compound nuclei fission a bit differently.
Further the densities of states in both cases are quite large ( both
systems being very deformed). Since most of the cross section at
sub-barrier energies is the two-neutron transfer one, we conclude that
the fissioning system is $^{240}$U with one unit of angular momentum added
to the high spin states populated.

\begin{figure*}
\includegraphics[width=13.5cm,height=12.cm]{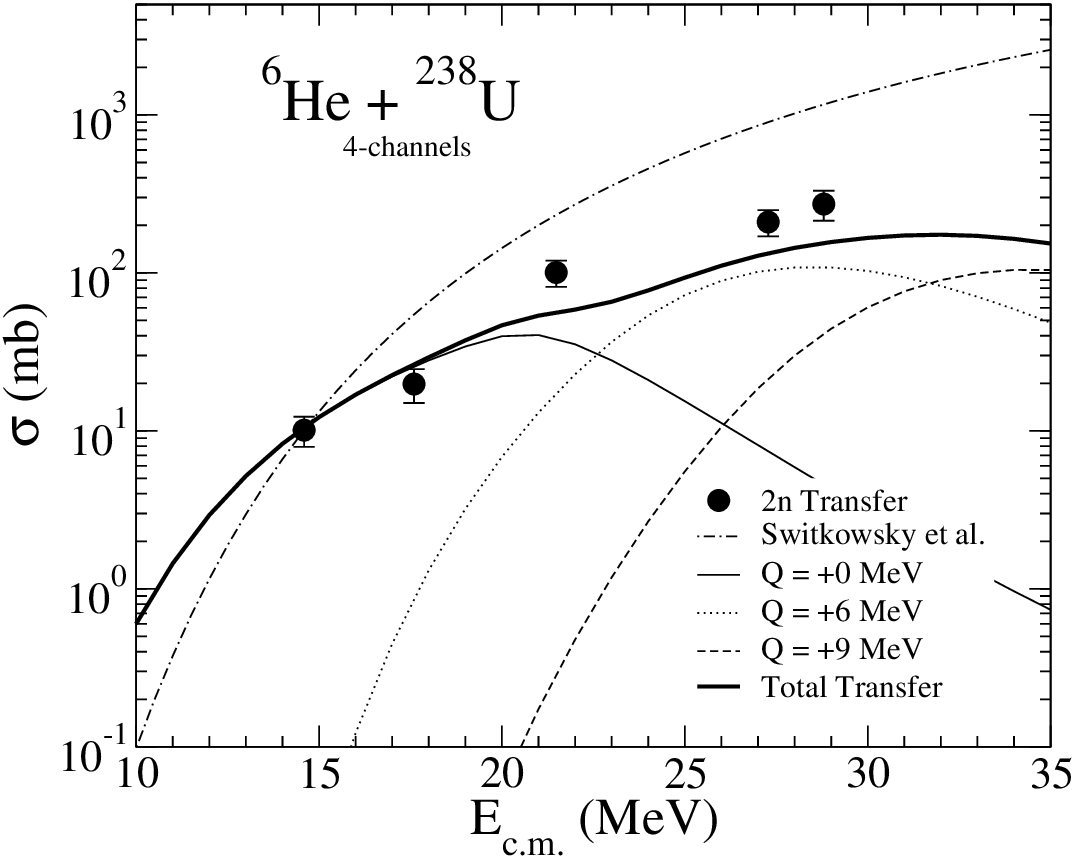}
\caption{\label{fig3} A figure caption.}
\end{figure*}

Switkowsky {\it et al.} \cite{Sw74} derived within the WKB approximation, a simple expression for the transfer cross section at energies well below the Coulomb barrier for the reaction $A_1 + A_2 \to A'_1 + A'_2$:
\begin{equation}
\sigma_T\approx\frac{1}{E_{\alpha}}\exp\left[ -4 \eta_{\alpha}\: {\rm arctg}\: \frac{\kappa}{k_{\alpha}}\right]
\label{swit}
\end{equation}
where $E_{\alpha}$, $\eta_{\alpha}$ and $k_{\alpha}$ refer to the incident channel ($\alpha =0$), and $\kappa$ is related to the binding energy of the transferred neutron, $\hbar \kappa=(2m_n B_{2n})^{1/2}$ and is equal to 0.307 fm$^{-1}$. Eq. (\ref{swit}) was obtained within the DWBA approximation for the transfer amplitude after employing the WKB form for the radial Coulomb wave functions which allows the use of the stationary point method. The position of the stationary point in the radial integral supplies the condition for the optimum Q-value, which comes out to be \cite{Sw74}
\begin{equation}
Q_{opt}=\left(\frac{Z'_1 Z'_2}{Z_1 Z_2} -1 \right)E_{\alpha} + \frac{Z'_1 Z'_2}{Z_1 Z_2} \frac{\hbar^2 \kappa^2}{2m_{\alpha}},
\label{qopt}
\end{equation}
if only low angular momenta are considred to contribute. The value of Q-optimal for the system $^6$He + $^{238}$U, where the first term in Eq. (\ref{qopt}) is identically zero comes to be 0.335 MeV. The results given by eq. (\ref{swit}) are compared with experiment for the 2n-transfer in the figure \ref{fig3} as the dashed-dotted curve. The cross section ,Eq. (\ref{swit}), was normalised to reproduce the datum at the lower energy where the approximation works best. It is clear that the transfer data at above barrier energies are overestimated. This calls for a more detailed consideration of two-neutron transfer with different Q-values at the higher energies. We turn to this in the following.

In the following, we describe a coupled channels calculation that takes into
account the fact the optimum Q-values change as the energy is lowered below the
barrier. In fact, within  the DWBA calculatiuon of \cite{Sw74}, appropriate at
sub-barrier energies, the optimum Q-value comes out to be about 0.335 MeV. At above barrier energies, the optimum Q-value could be much larger as larger
values of the angular momentum are involved. To simplify the discussion we
consider four channels: the entrance channel ($^6$He + $^{238}$U, $\alpha =0$), the two-neutron transfer channels ($^4$He + $^{240}$U) with Q = +0, +6, +9 MeV ($\alpha = 1,2,3$ respectively). For the transfer form factor we take $F(r) = F_0 \exp(-\kappa r)$ with $F_0 = 6$ MeV and $\kappa$ is related to the two neutron seperation energy. The coupled channel system is given by
\begin{widetext}
\begin{eqnarray}
-\frac{\hbar^2}{2\mu_{0}}\left[ \frac{d^2}{dr^2} - \frac{l(l+1)}{r^2}\right ]u_0(r)-(E_0-U_0(r))u_0(r)&=&\sum_{\alpha\ne0}F(r)u_{\alpha}(r)\label{3cc}\\
-\frac{\hbar^2}{2\mu_{\alpha}}\left[ \frac{d^2}{dr^2} - \frac{l(l+1)}{r^2}\right ]u_{\alpha}(r)-(E_{\alpha}-U_{\alpha}(r))u_{\alpha}(r)&=&F(r)u_0(r),\nonumber
\end{eqnarray}
\end{widetext}
The optical potencials  $U_i(r)$ are taken to be all equal to $U(r)$ of the Eq.
(\ref{op}).

We ignore the change in angular momentum in the centrifugal barriers, as the calculation of $<l>$ show. Since the two transferred neutrons populate different, orthogonal, states in $^{240}$U, the transfer cross section is the incoherent sum of the three contributions. The result is shown in figure \ref{fig3}, which clearly accounts well for the data. We have verified that the transfer coupling (with positive Q-valeus) has a very minor effect on the fusion cross section, supplying a less than 5\% reduction; part of the 40\% reduction used to normalize the fusion calculation of figure \ref{fig1}.  

In conclusion, we have considered several mechanisms to explain the large 2n-transfer cross-section at sub-barrier energies in the system $^6$He + $^{238}$U. It seems that a feasible mechanism is the incoherent contributions of two or more processes with quite different Q-values. Large two-neutron transfer cross section in the system $^6$He + $^{209}$Bi at energies near the Coulomb barrier has also been recently reported \cite{Yo05}. The mechanism may very well be similar to that reported in the present paper. Certainly, further work is required to elucidate the phenomenon, in particular the role of elastic breakup \cite{Yo05}.

\bigskip

{\bf Figure Captions}
\begin{itemize}
\item Figure 1: The full line is the fusion cross section from Eq. (\ref{sigfus}), multiplied by a factor 0.6, using an effective optical potential with the same parameters as in  ref. \cite{Ca03}. The dashed line is $^4$He fusion. The squares is the experimental data of \cite{Tr00} (solid squares) and \cite{Vi62} (open squares). 
\item  Figure 2: The 2n removal cross sections for $^6$He obtained with different approaches developed here. The full and dashed lines from Eqs. (\ref{sigtra}) and (\ref{sigtra1}) respectively.The experimental data are from ref. \cite{Ra04}. The vertical arrow indicates the position of the Coulomb barrier. 
\item Figure 3:
Transfer coupled channels calculations, from the Eq. (\ref{3cc}). See text for details.
\end{itemize}

\begin{acknowledgments}
We thank, T. Frederico, L. S. Ferriera, A.F.R. de Toledo Piza and M. Yamashita for helpful discussion and collaboration in the early stages of this paper. The first author is suppoted by FAPESP. The last two authors are supported partially by the CNPq, FAPESP and FAPERJ.
\end{acknowledgments}


\begin{references}
\bibitem{Ca05} For a review, see, L.F. Canto, P.R.S. Gomes, R. Donangelo and M.S. Hussein, ``Fusion and Breakup of Weakly Bound Nuclei", Phys. Reports, in press. (2005).
\bibitem{Tr00} M. Trotta, J.L. Sida, N. Alamanos, A. Andreyev, F. Auger, D.L. Balabanski, C. Borcea, N. Coulier, A. Drouart, D.J.C. Durand, G. Georgiev, A. Gillibert, J.D. Hinnefeld, M. Huyse, C. Jouanne, V. Lapoux, A. Lepine, A. Lumbroso, F. Marie, A. Musumarra, G. Neyens, S. Ottini, R. Raabe, S. Ternier, P. Van Duppen, K. Vyvey, C. Volant, R. Wolski, Phys. Rev. Lett. 84, 2342 (2000).
\bibitem{Ra04} R. Raabe, J.L. Sida, J.L. Charvet, N. Alamanos, C. Angulo, J.M. Casandjian, S. Courtin, A. Drouart, D.J.C. Durand, P. Figuera, A. Gillibert, S. Heinrich, C. Jouanne, V. Lapoux, A. Lepine-Szily, A. Musumarra, L. Nalpas, D. Pierroutsakou, M. Romoli, K. Rusek, M. Trotta, Nature(London) 431, 823 (2004).
\bibitem{Ca03} W.H.Z. Cardenas, L.F. Canto, N. Carlin, R. Donangelo, M.S. Hussein, Phys.Rev. C 68, 054614 (2003).
\bibitem{Ma78}  D.G. Madland and P.G. Young, Los Alamos Report No. LA 7533-mb (1978) (unpublished).
\bibitem{Al97} G.D. Alkhazov et al. Phys. Rev. Lett. 78 2313 (1997).
\bibitem{Da99} M. Dasgupta eta al. Phys. Rev. Lett. 82, 1395 (1999).
\bibitem{Hi02} D. Hinde et al. Phys. Rev. Lett. 89, 272701 (2002).
\bibitem{Ca02} W.H.Z. C\'ardenas, L.F. Canto, R. Donangelo, M.S. Hussein, J. Lubian, A. Romanelli, Nucl. Phys. A 703 (2002) 633.
\bibitem{Sw74} Z.E. Switkowsky, R.M. Wieland and A. Winther, Phys. Rev. Lett. 33 (1974) 840.
\bibitem{Pe38} R. Peierls, Phys. Rev. {\bf 54}, 918 (1938)
\bibitem{Ca98} L.F.Canto, R.Donangelo, L.M.de Matos, M.S.Hussein, P.Lotti, Phys.Rev. C58, 1107 (1998).
\bibitem{Yo05} P. A. De Young eta al. Phys. Rev. C 71, 051601(R) (2005).
\bibitem{Vi62}V.E. Viola and T. Sikkeland, Phys. Rev. 128, 767 (1962).
\end{references}
\end{document}